# Designing High-Performance and Thermally Feasible Multi-Chiplet Architectures enabled by Non-bendable Glass Interposer

HARSH SHARMA, and JANARDHAN RAO DOPPA, Washington State University, USA

UMIT Y. OGRAS, University of Wisconsin–Madison, USA

PARTHA PRATIM PANDE, Washington State University, USA

Multi-chiplet architectures enabled by glass interposer offer superior electrical performance, enable higher bus widths due to reduced crosstalk, and have lower capacitance in the redistribution layer than current silicon interposer-based systems. These advantages result in lower energy per bit, higher communication frequencies, and extended interconnect range. However, deformation of the package (warpage) in glass interposer-based systems becomes a critical challenge as system size increases, leading to severe mechanical stress and reliability concerns. Beyond a certain size, conventional packaging techniques fail to manage warpage effectively, necessitating new approaches to mitigate warpage induced bending with scalable performance for glass interposer based multi-chiplet systems. To address these inter-twined challenges, we propose a thermal-, warpage-, and performance-aware design framework that employs **architecture and packaging co-optimization**. The proposed framework disintegrates the surface and embedded chiplets to balance conflicting design objectives, ensuring optimal trade-offs between performance, power, and structural reliability. Our experiments demonstrate that optimized multi-chiplet architectures from our design framework achieve up to **64.7% performance improvement** and **40% power reduction** compared to traditional 2.5D systems to execute deep neural network workloads with lower fabrication costs.

Design Space Exploration; • Emerging Architectures; • System on Chip; • Machine Learning; • Interconnects.

## 1 INTRODUCTION

Emergence of 2.5D chiplet based architectures has drawn the attention of leading silicon manufacturers due to their superior power efficiency and lower fabrication costs [1] [2]. These 2.5D systems interconnect chiplets with a network-on-interposer (NoI) backbone through micro-bumps ($\mu$bumps). An interposer is an intermediate interface used in to facilitate electrical connections, allowing signals and power to be efficiently routed between chiplets. Interposer packaging approaches include silicon, organic substrate, and glass. Passive interposer, as the name suggests, support only interconnects. They are essentially large silicon chips fabricated with mature technology nodes with vertically routed through-silicon-via (TSV) links interconnecting the chiplets. Since the passive interposer constitutes 60–80% of the total 2.5D system area and remains largely inactive, they face challenges in meeting the future performance, cost, and reliability requirements, especially with the exponential growth in deep neural network (DNN) workloads [3].

Recent studies have compared different interposer materials at the package level [1] [4]. Glass interposer has gained attention in the microelectronics industry as a promising alternative to silicon-based interposer due to their superior electrical properties, tunable coefficient of thermal expansion (CTE), high mechanical rigidity, increased reticle limit, and significantly lower fabrication cost [1] [2] [5]. Unlike silicon interposer, glass interposer offers finer-line and spacing during fabrication, lower capacitance in the redistribution layer (RDL), and minimal crosstalk, enabling higher bus widths, lower energy per bit, higher communication frequencies, and extended interconnect range [2] [6]. These properties make them well-suited for achieving high performance and energy efficiency in 2.5D architectures. However, the thermal challenges are magnified with glass interposer, as their thermal conductivity ($TC_{glass}$ = 1x) is orders of magnitude lower than that of silicon (130x) and copper (400x). This results in thermal hotspots in glass interposer-based 2.5D architectures, limiting the achievable performance due to the inferior thermal profile.

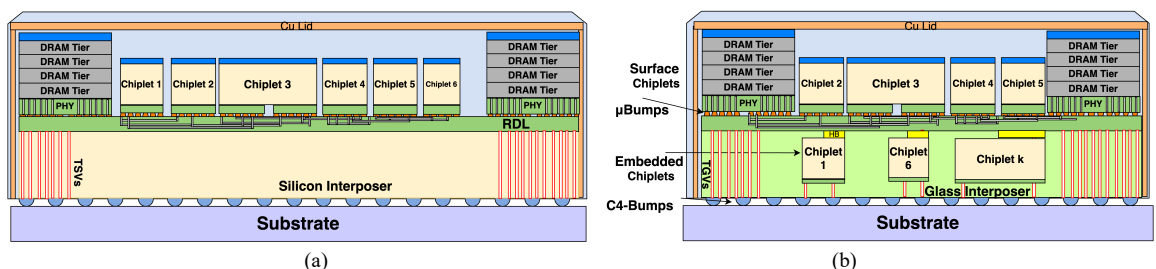

Figure 1: Illustrative example of (a) silicon- and (b) glass interposer systems (not to scale).

Another critical reliability concern in advanced packaging technologies is warpage, which refers to the abnormal bending or deformation of a multi-chiplet module (MCM) due to the mismatch in CTE between the chiplets and interposer. Warpage can be tolerated to a certain degree, but it becomes particularly pronounced with increasing system size. Excessive warpage can lead to chiplet cracking, substrate deformation, and misalignment issues, severely impacting manufacturability and yield. Key parameters affecting warpage include interposer thickness, material properties such as thermal conductivity and stiffness factor, operating temperature, chiplet placement, and effective module height [7]. Warpage is typically measured by assessing the vertical displacement at different points across a package. Beyond a certain size, conventional packaging approaches fail to effectively manage warpage, making large-scale integration of chiplets infeasible without additional structural optimizations. Multiple warpage mitigation techniques have been explored to minimize mechanical stress and improve MCM reliability using compensating layers in the RDL, flip-chip architectures, thermal buffers with higher CTE, and chiplet spacing adjustments [5]. While these techniques partially alleviate warpage-induced stress on the MCM, they either reduce the available silicon area and increase the complexity during chip fabrication or introduce additional processing overheads in the RDL layer. Warpage can also be reduced by increasing the stiffness of the material and lowering the effective height of the interposer package, distributing the mechanical stress evenly across the structure. A stiffer material (with higher Young's modulus) increases resistance to bending, while a lower effective height reduces the effect of the force due to the lever effects, which increases warpage. This can be achieved by doping existing interposer materials to enhance their mechanical properties or introducing alternative materials with inherently higher stiffness and lower coefficient of thermal expansion, thereby reducing height and introducing stiffness.

Warpage with glass interposer is an order of magnitude higher than silicon interposer [5]. Even though the fabrication cost of glass is much lower compared to silicon-based interposer, warpage makes it impractical to create glass interposer-based 2.5D systems above a certain area threshold. Embedding functional components within the glass interposer has been explored as a technique to reduce the amount of warpage [1]. For instance, silicon interposer has a Young's modulus of 130-180 giga-pascals (GPa), while glass interposer has a much lower Young's modulus of around 50-75 GPa. By embedding chiplets within the interposer, this approach not only reduces the interposer height—thereby lowering the moment arm that contributes to warpage—but also increases the overall stiffness of the structure. Fig. 1(a) & (b) illustrate a 2.5D silicon interposer with surface chiplets and a glass interposer system with both surface and embedded chiplets, respectively. Embedding logic in silicon interposer has been widely discarded due to electrical crosstalk challenges, but it is feasible for chiplet systems with glass interposer as glass is electrically an insulator [1]. By embedding more chiplets within an interposer with a smaller surface area, a glass interposer system can achieve higher compute capability than a silicon-based 2.5D system with equivalent or lower fabrication costs.

Prior works have investigated integrating memory, logic, and inter-chiplet communication routers to transform the interposer from an interconnection medium to more functional/logic components [1] [5]. Signal integrity, manufacturability, energy consumption, frequency of operations, size, and thermal cycling profiles of 2.5 D systems have been investigated and optimized as separate objectives, optimizing each metric in isolation [2] [5]. However, *no prior work* has holistically examined the entire design space, encompassing material impact, chiplet placement, warpage constraints, and system-level performance and thermal considerations to design a multi-chiplet system. Such a holistic multi-objective optimization approach is crucial, as optimizing one design objective can dramatically affect others. For instance, selecting a particular NoI topology and corresponding chiplet placement influence achievable latency, energy consumption, and hence the peak temperature. This, in turn, affects mechanical stress, degree of warpage, overall performance, and MCM reliability. Similarly, due to its high TC, silicon-based chiplet systems are more effective in dissipating the heat generated by the surface-mounted chiplets than glass interposer-based 2.5D systems. Significantly lower TC of glass impedes heat dissipation and eventually causes thermal hotspots in the system. Hence, chiplets with relatively lower power consumption must be employed when incorporating glass as the interposer material. To balance power, performance, and warpage constraints effectively, we adopt an end-to-end design framework that considers all levels of the design space. We must improve the overall performance, energy efficiency, and reliability of an MCM system without introducing additional fabrication costs.

In this work, we propose a thermal-, warpage-, and performance-aware design and optimization framework that employs architecture and packaging co-optimization to disintegrate the 2.5 D system suitably between surface and embedded chiplets based on the NoI architecture and the type of interposer. Our design optimization methodology enables high-performance and energy-efficient MCM design specifically targeted for DNN workloads by efficiently partitioning between surface and embedded chiplets, mitigating thermal hotspots and structural deformation while ensuring system scalability. We consider various types of processing-in-memory (PIM) chiplets due to their suitability in accelerating DNN inferencing workloads. We achieve this by using a sample-efficient Bayesian optimization technique to explore huge design space to balance multiple conflicting design objectives. This allows optimized 2.5D architectures to achieve high performance and power efficiency without excessive warpage or thermal hotspots.

It is essential to establish suitable performance, MCM warpage, and thermal trade-offs in chiplet based systems depending on multiple parameters. Relevant parameters include the system size, the type and the number of corresponding chiplets, power density, and the NoI architecture. *To the best of our knowledge, this is one of the initial investigations that establishes the necessary design trade-offs for a multi-chiplet glass interposer system*. We demonstrate high performance and energy efficiency without introducing thermal hotspots compared to traditional 2.5D systems to execute server-scale DNN workloads. Moreover, the design framework enables multi-chiplet architectures with lower fabrication costs with the same or improved performance. The major contributions of this work are:

- We propose a design optimization framework using wide-body chiplet systems enabled by glass interposer targeting large-scale DNN workloads. We integrate different types of PIM chiplets to improve performance, energy and thermal efficiency, and fabrication cost compared to silicon interposer-based 2.5D counterparts.
- Comprehensive experimental evaluations with server-scale DNN workloads demonstrate that our optimized heterogenous multi-chiplet architecture with glass interposer achieves up to 64.7% and 60% reduction in latency and energy with up to 3.53x lower fabrication cost than silicon interposer-based 2.5D counterparts [8].

The rest of the paper is organized as follows. Section 2 describes relevant prior work. Section 3 presents the design principles and the proposed optimization framework. Section 4 presents the detailed experimental results and analysis. Finally, Section 5 concludes the paper by highlighting the salient contributions.

## 2 RELATED WORK

This paper focuses on designing chiplet-based architectures with glass interposer. Hence, we review the related work in two parts: chiplet-based manycore architectures and previous research with interposer as packaging technology.

**Chiplet-based Architectures:** Both application-specific and general-purpose 2.5D chiplet architectures have been explored in the literature. All existing architectures were primarily designed with passive interposer. Design space exploration of 2.5D-based systems considering technology nodes, chiplet-sizes, big-little chiplet paradigm, and multi-link network frequency architectures have been studied [9] [8] [10] [11]. The NoI paradigm becomes crucial as the communication demand increases when many chiplets are integrated on the same substrate [11]. Multiple NoI architectures have been proposed in the literature. However, most of these architectures are based on conventional multi-hop interconnection architectures, such as mesh or torus [11] [10] [12]. SWAP is a server-scale application-specific 2.5D architecture proposed for DNN inferencing workloads [9]. The SIAM framework enables fast design space exploration of 2.5D-based systems [11]. SIMBA introduces tiling optimizations on a fixed 2D-mesh NoI topology for executing DNN workloads [12]. Recent work discusses the advantages of integrating heterogeneous chiplets on the interposer for high performance with reduced fabrication cost [2] [4]. HexaMesh is a compact-packing high-fan-out chiplet-based interconnection architecture that improves bisection bandwidth and has inherently larger router ports with star-like connections to its possible neighbors [13]. Recently, a hexagonal topology called FoldedHexaTorus is proposed, which principally introduces skip connections on the HexaMesh topology [14]. Hence, the radix of the NoI, the number of router ports, and chiplet placement structure remains identical. Floret is a space-filling-curve based NoI architecture for 2.5D systems that accelerates CNN workloads by employing a dataflow-aware mapping along the contiguous chiplets [8]. The Kite family of NoI topologies for 2.5D systems are primarily Torus-based [10]. Passive interposer has become a mainstream solution to address the bandwidth demands of emerging applications. Recently, heterogeneous multi-chiplet architectures for accelerating large-language models have also been proposed [15]. AMD has recently introduced a heterogeneous line of chiplet platforms called MI-300 to enable IP reusability and to co-optimize performance, energy consumption, and fabrication costs [3]. In addition, 3D-chiplet architectures have been demonstrated to maximize bandwidth [2] [4] [16]. This wide variation in physical realization requires a thorough design space exploration where the advantages and limitations of individual chiplet architecture are leveraged suitably.

**Interposer as packaging technology:** IntAct is one of the earliest architectures demonstrating low latency interconnects on a chiplet-based system [4]. It is a 6-chiplet 96-core architecture with routing logic and peripheral test circuitry such as JTAG implemented within the interposer. Active interposer-based packaging methods including Intel EMIB, TSMC InFO-oS, and IntAct, enable limited logic embedding in silicon interposers [4] [17] [18] [19]. However, these designs incorporate only lightweight logic blocks such as routers, buffers, or small interconnect units but not full chiplets. Moreover, silicon-based interposer with embedded logic faces challenges in meeting the future performance, cost, and reliability requirements. It has been demonstrated that silicon interposer is not cost-effective, and interconnects tend to be lossy if chiplets are physically far apart, exacerbating performance degradation in a wide-body system [1] [2]. Multiple studies demonstrated signal integrity, electrical data transfer improvement, and stress tests for glass-based interposer [16]. It has been shown that glass interposer can avoid thermal coupling/cross-talk among chiplets. Recently, embedding memory has been demonstrated within the glass interposer [2] [16]. *In this paper, we bridge this gap in the current state of knowledge*

Table 1: Different chiplet configurations along with their total storage, area, TOPS, and Energy/MAC considered in this work.

| Type | Tech Node | Type | Size | Tech | Bits/cell | ADC precision | Frequency | Total Storage (KB) | Area ($mm^2$) | TOPS | Energy(J)/MAC |
|---|---|---|---|---|---|---|---|---|---|---|---|
| 1 | 22 | Standard [22] | 128x128 | ReRAM | 2 | 8 | 100MHz | 1196 | 4 | 30 | .87E-12 |
| 2 | 22 | Shared_ADC [23] | 768x768 | SRAM | 1 | 8 | 100MHz | 1080 | 8 | 27 | .30E-12 |
| 3 | 22 | Adder [21] | 64x64 | SRAM | 1 | 8 | 100MHz | 108 | 4 | 11 | .18E-12 |
| 4 | 22 | Accumulator [24] [25] | 256x256 | ReRAM | 2 | 8 | 100MHz | 2400 | 4 | 35 | .22E-12 |
| 5 | 22 | ADCLess [26] | 128x128 | SRAM | 1 | 1 | 100MHz | 300 | 4 | 3.8 | .27E-12 |

*by proposing design principles and efficient optimization methods to enable heterogeneous integration of multi-chiplet architectures using glass interposer.*

## 3 ARCHITECTURE-PACKAGING CO-OPTIMIZATION FOR CHIPLET DISINTEGRATION

Integrating chiplets on interposer presents unique challenges and opportunities in terms of overall performance, system storage capacity, energy consumption, warpage, and thermal efficiency. This section presents the salient features of the proposed design framework that enables disintegration between chiplets to achieve high performance without creating any thermal hotspots. We specifically consider processing-in-memory (PIM)-based chiplets as the computing platform for accelerating DNN workloads [9]. Table 1 shows different device types, storage capacity, frequency of operation, area footprint, $TOPS$ and $Energy/MAC$ for the different PIM chiplets considered in this work. As each PIM chiplet area may vary, the total number of chiplets, depending on the chosen types within the given interposer area, will also vary. Moreover, the characteristics of silicon and glass interposer vary in terms of thermal conductivity and warpage.

### 3.1 Background and Preliminaries

**Silicon vs. Glass Interposer**: With its higher TC, a silicon interposer is effective for heat dissipation. However, due to the defect density on wafers, there is a scaling limitation, and the interposer dimensions must be within a reticle area limit. Increasing the compute capability of either chiplets or interposer is one way to provide higher performance using the same system size. However, current manufacturing processes do not allow for the embedding of chiplets except for small routers and repeaters (e.g., EMIB bridges) in the silicon interposer due to limited signal integrity arising from cross-talk over long distance [1] [2]. Glass interposer, on the other hand, offer the capability to embed PIM chiplets. This flexibility has the potential to improve the trade-offs in design objectives by balancing the warpage and thermal profile.

Furthermore, as highlighted in recent studies, developing thermal vias and specialized cooling techniques with copper bridges have further enhanced the thermal performance of glass interposer, making them a viable solution [16]. Considering the signal integrity, the reduced parasitic capacitance resulting from a glass interposer's lower dielectric constant minimizes power loss and delays in signal transmission, offering an energy-efficient alternative to silicon, especially in high-speed applications. In the glass interposer-based systems considered in this work, the most significant advantage comes from embedding PIM chiplets, reducing the distance to the DRAM module (3D stacked on the surface) compared to silicon interposer-based 2.5D counterparts. This proximity is crucial for high-speed memory access, leading to shorter interconnect lengths, hence, lower latency and higher bandwidth [20]. The lower dielectric constant of the RDL layer over glass interposer also contributes to improved signal integrity and faster signal propagation, enhancing the communication between the processor and memory. However, using a glass interposer introduces a critical thermal bottleneck. Given the 130x lower TC of glass than silicon, the heat generated by chiplets must travel through a resistive path to the heat sink. As the chiplets are connected directly to the interposer through microbumps, glass restricts the heat flow, causing localized heating and, consequently, increased peak temperature $T_{peak}$. Exceeding the thermal budget can result in decreased reliability and performance degradation. Moreover, embedding more chiplets will increase power density, exacerbating the thermal limit. Consequently, in the proposed design framework, we balance the number and types of both surface and embedded chiplets to optimize the thermal profile while balancing warpage, energy consumption, and system performance.

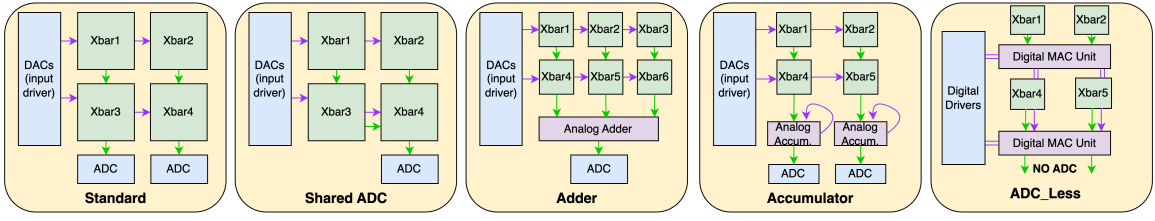

Figure 2: Different PIM chiplet configurations considered in our multi-chiplet system design space exploration to optimize multiple design objectives for both silicon- and glass interposer-based systems (not drawn to scale).

**Heterogenous Chiplet Architectures:** As chiplet activity directly impacts the thermal profile, it is important to consider their individual characteristics and interaction with the interposer. Moreover, the package-architecture interaction dictates the system warpage, thermal profile, and achievable performance. Fig. 2 illustrates the top-level organization of the different PIM chiplet configurations considered in our multi-chiplet system design space exploration to accelerate DNN workloads. Each chiplet consists of multiple processing elements (PE), where each PE has $T$ tiles and a PE buffer. Each tile has multiple PIM crossbars, where each crossbar unit is an SRAM or ReRAM cell. Each cell adopts $b$ bits/cell resolution. The multiply-and-accumulate (MAC) operation in ReRAM is performed by applying analog row inputs to the stored weights on the ReRAM cell, generating and accumulating currents along the column wires. The accumulated current is then read out via the output analog-to-digital converter (ADC). In case of SRAM PIM, the MAC operation is performed by applying multiple read word-line pulses as digital row inputs to the stored weights in SRAM cells. The bitwise MAC results are generated through binary-weighted capacitor-based charge sharing and are accumulated as voltage differences along the bit-lines. The accumulated voltage is then digitized using a shared ADC circuit or an adder-based accumulation unit [21]. Each tile has necessary buffers to store intermediate activations during the MAC operations, and peripheral circuits to implement nonlinear functions such as rectified or Gaussian error linear unit (ReLU/GeLU), Sigmoid, etc. This work considers the following PIM chiplets (refer to Table 1) as exploratory examples:

- *Standard PIM configuration*: A standard configuration has a square crossbar array (128X128 is most common) with ADC on each column [22]. Both SRAM and ReRAM can be used to implement the MAC operations. We need a digital-to-analog converter (DAC) for each input, and the DNN weights are loaded onto the crossbar arrays. The MAC operation occurs through the crossbars, which are read through the output ADC. We consider a ReRAM-based PIM with 128X128 crossbars in this work. This configuration is referred to as "*Standard*" throughout this manuscript.
- *Shared ADC configuration*: In this configuration, the outputs are reused across different rows and columns. An SRAM-based PIM with shared ADC has been demonstrated [23]. In this case, analog outputs across different columns are summed up on the wires. The energy consumption is lower because the ADC is shared among multiple columns. Output reuse between columns decreases ADC energy but gives rise to lower input reuse and higher DAC energy. We refer to this design as "*Shared*" in the manuscript.
- *Adder configuration*: In this case, the crossbar structure is rectangular. It activates a smaller subset of the crossbar during the operation. We consider an SRAM-based PIM with this configuration [21]. This configuration uses an analog adder at the output before the ADC. The ADC reads the resulting sum rather than each output individually, hence amortizing the energy. We refer to this design as "*Adder*" in the manuscript.

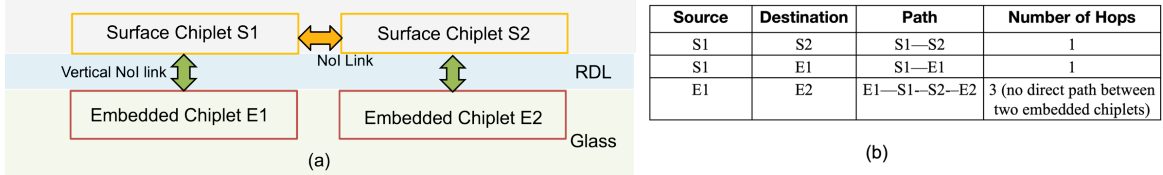

Figure 3(a): Illustrative example of inter-chiplet communication between surface and embedded chiplets in multi-chiplet system design space (not drawn to scale); (b) Communication hops along with the path between two chiplets depending on their z-location.

- *Analog accumulator configuration*: This configuration employs an analog accumulator at the output before the ADC to *reuse outputs across different cycles,* significantly reducing the energy consumption. We consider a ReRAM-based PIM in this configuration [24] [25] and refer to it as "*Accumulator*" in the manuscript.
- *ADC-less PIM configuration*: In this case, we use the crossbar array for weight storage, and the computations are performed with digital MAC units. It is primarily a processing-near-memory paradigm. We consider SRAM as a memory array in this configuration [26]. We refer to this configuration as "$ADC_{less}$" in the manuscript.

**Surface and Embedded Chiplets:** Principally, irrespective of the chiplet type on the device level, ReRAM cells (generally 1T-1R) are much smaller than SRAM-based cells (6T). Hence, a ReRAM-based chiplet has higher storage and compute capability and power density due to its smaller area footprint. Moreover, the operational temperature (dictated by the heat flow, which is suboptimal in the case of glass interposer) of PIM chiplets significantly impacts their reliability and achievable performance, particularly through noise-induced variations in DNN weight values [27]. ReRAM cells store weights using conductance, which varies with temperature [28]. Any thermal fluctuations within the ReRAM cells can introduce noise, potentially resulting in erroneous computations and compromising the inference accuracy of the underlying DNN model [29] [30]. For ReRAM-based chiplets, the conductance variation ($\sigma(T)$) follows a Gaussian thermal noise relationship [31]. At standard room temperature (300K) with typical ReRAM conductance (*G*) of 100 micro-siemens (*μS*), this results in conductance variation of approximately 0.1%. However, at elevated temperatures (e.g., 350K), the temperature-induced variation can increase by up to 8%, potentially affecting the weight precision and, consequently, the DNN inferencing accuracy.

Hence, placing multiple ReRAM chiplets in proximity on glass interposer is challenging, as higher operating temperatures degrade the DNN accuracy due to localized heating. On the other hand, SRAM is a mature technology to design PIM chiplets [32]. SRAM chiplets have larger footprint than their ReRAM counterparts for the same level of processing capabilities; hence, they have comparatively lower power density. In the case of Shared_ADC, as the ADC is shared among multiple PEs, the power density is much lower. However, as SRAM chiplets are comparatively larger, their effect on the warpage is more pronounced and should be placed closer to the center of the chip. Hence, there is an intricate dependency of chiplet placement and multi-chiplet configuration on achievable performance, warpage, and thermal profile.

**Table 2: Average number of NoI router ports for a standard NoI**

| System Size = 100 | Kite | SIAM | HexaMesh | SWAP | Floret |
|---|---|---|---|---|---|
| Avg. # of router ports | 4 | 3.6 | 4.86 | 2.8 | 2.1 |

*Since embedding ReRAM chiplets is prohibitive due to the thermal challenges, we only embed SRAM-type chiplets*. The advantage of more efficient inter-chiplet communication due to lower capacitance and energy per bit is balanced with the thermal limitations. Moreover, the NoI governs the inter-chiplet data exchange, hence the communication latency and power consumption. Below, we discuss in detail the impact of NoI architecture in glass-based 2.5D systems.

**NoI Design Strategy:** NoI is the communication backbone, enabling the integration of multiple chiplets in a 2.5D system. It governs inter-chiplet data exchange latency and power consumption. Each chiplet is associated with a router and corresponding links. The number of NoI links associated with each chiplet depends on the interconnection architecture. In this work, we consider the following four recently proposed NoI architectures: **1)** 2D-Mesh NoI connecting chiplets in a grid [11]. SIAM, SIMBA, and MI-300 are previous chiplet explorations that employ a grid-like 2D-Mesh architecture, which we refer to as 'Mesh' [3] [12]. **2)** Kite is a torus-based NoI, which introduces skip connections [10]. **3)** HexaMesh is a concentrated interconnect architecture with a star-like topology between neighboring chiplets [13][33]. 'HexaMesh' can be seen as a 3D-Mesh projected onto a 2-dimensional space, with each intermediate chiplet connected to six neighbors, hence the name 'HexaMesh'. **4)** Floret is a space-filling curve (SFC)-based hierarchical NoI architecture where each intermediate chiplet is at least connected to the previous and the next chiplet [8]. Table 2 shows the average number of router ports associated with each NoI architecture considered here for a 100 chiplet system placed in a regular 10X10 grid. It is evident that Floret has least number of ports and inter-router links, and HexaMesh has the most.

Fig. 3(a) shows an illustrative example of inter-chiplet communication in glass-based chiplet system. As we embed a chiplet, it is connected directly to the surface chiplet and has a vertical link between the surface and embedded chiplets. Fig. 3 (b) shows the source, destination, the shortest path necessary between different pairs of chiplets, and the number of hops. As there is no direct path between the two embedded chiplets, communication must be facilitated through the surface NoI topology.

### 3.2 Problem Setup and Multi-Objective Optimization Formulation

We consider a multi-chiplet system with $N$ PIM-based chiplets distributed over an interposer of area $A$, where each chiplet is placed on the interposer [3] [12] [16]. Each chiplet may be among the above five choices (refer to Table 1). It should be noted that the tera-operations-per-second (TOPS denoted by $\beta$), energy consumption per MAC operation (denoted by $\Gamma$), peak power consumption ($P$), and the amount of storage capacity ($S$) for these five types of chiplets vary.

Our goal is to jointly optimize the overall performance and energy efficiency within the thermal constraints by selecting the composition of heterogeneous chiplets and their placements along the interposer and mapping neural layers to chiplets to co-optimize the trade-offs. We keep the allowable peak temperature within 75 °C ($T_{max}$). If the temperature exceeds $T_{max}$, performance degradation occurs during the data loading phase from DRAM chiplets [34] [20]. ReRAM-based chiplets experience conductance drift due to thermal variations, impacting the predictive accuracy of DNN models. As the temperature increases, the OFF-state conductance of ReRAM cells increases exponentially, and the noise margin reduces [28] [35]. This impacts the usability of ReRAM- chiplets depending on the overall thermal profile of the system and the sensitivity of DNN weights to thermal-induced variations [36].

Without loss of generality, each DNN workload can be characterized by the properties of different neural layers. A DNN layer is characterized by its corresponding kernel size, input/output features (which jointly dictate the number of activations and storage requirements for each layer), and layer sparsity. DNN inferencing requires high-precision computation of activations for each neural layer. This process requires many MAC operations, influencing the choice of PIM chiplets. A neural layer's computation and storage requirements depend on its characteristics. We aim to minimize the total energy consumption based on the type of chiplets chosen to execute a particular layer with the lowest latency. To balance multiple conflicting design objectives, we formulate a *multi-objective optimization (MOO) problem* to find the composition and suitable placement of heterogenous chiplets and associated neural layer to chiplet mapping that achieves the best trade-offs between latency and energy consumption without violating the specified thermal constraint.

Table 3: Summary of notations used in this article.

| | | | | | |
|---|---|---|---|---|---|
| $A$ | Total interposer area | $d_c$ | Candidate placement of chiplets | $Q$ | Inter-chiplet connectivity |
| $W$ | Neural layer storage requirement | $k$ | Number of DNNs | $h$ | Number of hops |
| $m$ | Neural layer compute requirement | $\pi_{l \to c}$ | Neural layer to chiplet mapping policy | $\delta$ | Router delay |
| Act | Neural layer activations | $\Delta \psi$ | Coefficient of thermal expansion (CTE) | $E_{link}$ | Energy consumed per bit per hop |
| $s$ | Neural layer sparsity | $\lambda$ | Thermal conductivity | $E_{router}$ | Energy consumed per bit per intermediate router |
| $S$ | Neural layer sensitivity | $D$ | Stiffness Factor | | |
| $t$ | Type of chiplet | $k$ | Youngs' modulus | $K_B$ | Boltzmann constant |
| $\alpha$ | Individual count of chiplets | $\kappa$ | Warpage | $G$ | Conductance Value |
| $\beta$ | Tera-operations-per-second (TOPS) | $\Gamma$ | Energy(J)-per-MAC | $\sigma$ | Conductance variation |
| $T_{max}$ | Allowable Peak temperature | $S$ | Chiplet storage capacity | $\theta$ | DNN parameters |

**I.      Inputs:**

The inputs to the optimization framework are as follows: the type of interposer, the total interposer area $A$; the DNN workload characteristics (e.g., weights storage requirement $W_l$, compute requirement $m_l$, activations $Act_l$ for each neural layer $l$). The different types of chiplets with their individual count $\alpha = [\alpha_1, \alpha_2, ..., \alpha_t]$, where $\alpha_1, \alpha_2,... \alpha_t$ are the number of occurrences of $t$ types of chiplets. For a system with embedded chiplets, the total area available is the sum of surface and embedded area ($A = A_{surface} + A_{embedded}$). There may be multiple combinations of chiplets that satisfy the overall requirement. However, as the characteristics of each chiplet in terms of TOPS, storage capacity, and area vary, we optimize over combinatorial multi-chiplet configurations within the peak thermal limit $T_{max}$. As the thermal characteristics between glass and silicon vary, the optimal heterogeneous chiplet configuration varies. The system configuration should be such that it maximizes overall TOPS (a proxy for performance) and storage capacity with minimum energy consumption when all the chiplets are fully utilized without introducing thermal hotspots. For the DNN workloads (shown in Table 5) considered in this paper, we use the same optimized heterogeneous system configuration to evaluate the overall performance for all the workloads. Table 3 summarizes all the symbols used in this work.

**II.     Design Variables**

There are three types of design variables for the multi-chiplet design $d = (\alpha, d_c, \pi_{l \rightarrow c})$. $\alpha = [\alpha_1, \alpha_2, ..., \alpha_t]$ denotes the individual count of different types of chiplets (composition of heterogeneous chiplets), where $\alpha_1, \alpha_2,... \alpha_t$ are the number of occurrences of $t$ types of chiplets; $d_c$ corresponds to the candidate placement of the chiplets conditioned on the composition of heterogeneous chiplets $\alpha$. $\pi_{l \rightarrow c}$ corresponds to neural layer ($l$) to chiplet ($c$) mapping.

**III.    Design Objectives**

Next, we explain the evaluation of the design objectives, which are divided into two parts: ***Architectural design objectives*** and ***Package-level design objectives***. Architectural design objectives include latency and energy consumption. Package-level design objectives/constraints include warpage and thermal profile.

***Latency***: We evaluate the end-to-end latency incurred in DNN inferencing for a candidate design $d$ (with latency $Lat(d)$) considering individual chiplet characteristics such as TOPS and the average hop-count among communicating chiplets. For any DNN workload considered here, the latency is dictated by the slowest compute and communication stage, which further depends on the storage capacity, and available TOPS of each chiplet. We consider DNNs with different neural layer architectures—including linear (e.g., VGG), residual (e.g., ResNet, MobileNetV2), and dense (e.g., DenseNet) connections for performing inference tasks when designing a chiplet-based system. However, mapping different DNNs dynamically to a chiplet-based system is challenging. The common property of DNN inferencing is that the activations flow from the $i^{th}$ layer to the $(i + 1)^{th}$ layer. Maintaining contiguity on the physical NoI layer between two consecutive neural layers reduces communication overhead. We assume that each chiplet has sufficient buffers to store the intermediate activations associated with the skip connections, which flow through the same NoI links. Hence, it is imperative to map communicating DNN layers to neighboring chiplets. Moreover, as the storage requirement of each DNN layer varies, a neural layer could either fit within a single chiplet or get distributed over multiple chiplets. The forward pass involves computing activations ($Act$), which necessitates PIM chiplets to have ample storage [37]. However, if the storage requirement of a neural layer is relatively large (common for later layers in a DNN to be bigger), then the layer is spread onto multiple chiplets. Hence, the inter-chiplet traffic pattern increases, which exacerbates the communication latency. If the TOPS of a given chiplet is higher, the compute is faster. Hence, the overall latency is a balance between the computation and communication time, which depends on the available storage and compute capacity of an individual chiplet.

For $k$ individual DNNs mapped on to $N$ chiplets, the latency can be estimated as:

$$Lat(d) = max(Compute_{Lat(d)}, Comm_{Lat(d)}) \quad (1)$$

For $k$ DNNs with $L_1, L_2..,L_k$ number of neural layers mapped on the chiplet of type $t$ with TOPS $\beta_t$ (the TOPS for the five different chiplets are given in the Table), the compute latency can be approximated as:

$$Compute_{Lat(d)} = \sum_{j=1}^{k} \sum_{l=1}^{L_j} \left(\frac{m_{lj_t}}{\beta_t}\right) \quad (2)$$

where $m_{lj_t}$ is defined as the number of operations for neural layer $l$ of the $j^{th}$ DNN mapped on a chiplet of type $t$. The communication latency for the chiplet-based system with inter-chiplet connectivity $Q_{ij}$ (1 if a direct communication link exists between $i^{th}$ & $j^{th}$ chiplet else 0) can be estimated as:

$$Comm_{Lat(d)} = \sum_{j=1}^{N} \sum_{i=j}^{N} Q_{ij} \cdot Act_{ij} \cdot \left(h_{ij} + \delta_{router_j}\right) \quad (3)$$

Where $h_{ij}$ represents the number of hops the data packet passes through between $i^{th}$ and $j^{th}$ chiplet; and $\delta_{router_j}$ is the $j^{th}$ router delay in the source-to-destination path. The total latency is calculated by considering both the compute and communicate latency for each design $d$.

***Energy***: For $k$ individual DNNs mapped to $N$ chiplets with, the total energy consumption can be estimated as:

$$E(d) = (Compute_{E(d)} + Comm_{E(d)}) \quad (4)$$

For $k$ DNNs with $L_1, L_2..,L_k$ number of neural layers, the compute energy consumption of layer $l$ from $j^{th}$ DNN depends on its size, mapped chiplet of type $t$ with energy/MAC $\Gamma_t$, its sparsity $s_{lj}$, and the corresponding number of MAC operations $m_{lj}$ needed to execute it as shown:

$$Compute_{E(d)} = \sum_{j=1}^{k} \sum_{l=1}^{L_j} s_{lj} * m_{lj} * \Gamma_t \quad (5)$$

As the energy consumption per MAC differs for each PIM chiplet, the total DNN energy consumption depends on which chiplets are used for mapping different DNN layers. The communication energy for the chiplet-based system with inter-chiplet connectivity $Q_{ij}$ can be estimated as:

$$Comm_{E(d)} = \sum_{j=1}^{N} \sum_{i=j}^{N} Q_{ij} \cdot Act_{ij} \cdot \left(h_{ij} \cdot (E_{link} + E_{router})\right) \quad (6)$$

where $h_{ij}$ is the number of hops the data travels between chiplet $i$ and chiplet $j$, $E_{link}$ is the energy consumed per bit per hop. Depending on vertical or horizontal link, $E_{link}$ varies [38]. $E_{router}$ is the energy consumed per bit per intermediate router the data packet passes through in the source-to-destination path. As the energy per MAC differs for each PIM chiplet, the total DNN energy consumption depends on which chiplets are used for mapping different DNN layers and the communication energy depends on the NoI backbone.

**Fabrication Cost:** The thermal profiles and performance characteristics determine which type of chiplet should be embedded. If the number of embedded chiplets decreases, we will have more surface-mounted chiplets. This will increase the lateral area of the system, leading to a larger interposer area and, hence, higher fabrication costs. For a chiplet-based system, the normalized fabrication cost of chiplets is expressed as:

$$C_{system} = \frac{L_{ref}}{L} \times e^{-D_0(A_{ref} - A_{system})} \quad (7)$$

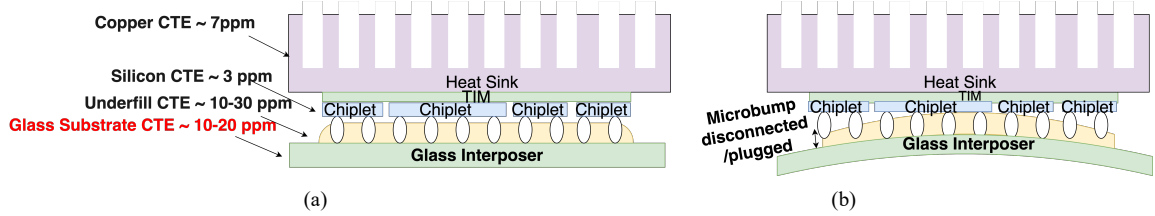

Figure 4: Illustrative example of warpage effecting the chiplet system (not drawn to scale); (a) Expected system; (b) Warpage disconnecting connections for chiplets on the edge.

where $L_{ref}$ is the number of chiplets per wafer in the reference system, and $L$ is the number of chiplets per wafer for the system under consideration. $D_0$ represents the defect density, and $A_{ref}$ is the NoI area of the reference system. We consider a 2.5D system with $864\ mm^2$ area as the reference in this work [39]. The system area consists of the area of the chiplets, the associated TSVs for each NoI link, and the interposer. The relative fabrication cost of two different systems is [11]:

$$C_{relative} = e^{-D_0(A_{System_1} - A_{System_2})} \qquad (8)$$

The relative fabrication cost primarily depends on the differences in the total system areas, which is computed as:

$$C_{system} = C_{interposer} + C_{chiplets} + C_{TSVs} \qquad (9)$$

The NoI area is influenced by factors such as the interposer area, the size of the routers, and the number of inter-chiplet links. Additionally, the system cost depends on both the type of chiplets used and their individual count. Since a glass interposer is approximately eight times (8x) cheaper than a silicon interposer, the number of chiplets in the system can be increased without significantly raising the cost. However, TSVs (through-silicon vias) in glass interposer, often referred to as TGVs (through-glass vias), are typically thicker and have a larger diameter compared to those in silicon interposer, all within the same footprint. This difference in TSV characteristics leads to distinct cost overheads. Table 4 provides a detailed comparison of the TSV costs between silicon and glass interposer-based system.

### IV. Constraints

We have three package-level constraints: one for interposer area, one for peak temperature, and one for warpage.

**Area constraint.** The total chiplet area is constrained by the given interposer area limit $A$ as:

$$\sum_{i=1}^{k} a_i \cdot \alpha_i < A \qquad (10)$$

where $a_i$ represents the area of the $i^{th}$ chiplet. This means that we only consider the heterogeneous chiplet compositions $\alpha = [\alpha_1, \alpha_2, \ldots, \alpha_t]$ that meet this area constraint during the design space exploration.

**Temperature constraint.** We constrain the peak temperature of the system to not exceed $T_{max}$. We employ $T_{max}=75$ °C, noting that our MOO formulation and optimization methodology is general.

**Warpage constraint**: Warpage is a major reliability concern in advanced packages that refers to abnormal bending of shape due to the mismatch in the CTE of different materials. Ensuring warpage within a tolerable limit is an important package-level design metric. Fig. 4 shows an illustrative example of warpage in a glass interposer-based chiplet system. Warpage can be measured by the variation in vertical height at different positions on the interposer package. The warpage at a distance $x$ from the center of the interposer is given by [7]:

$$\kappa(x) = \frac{\tau \cdot \Delta\psi \cdot \Delta T}{2 \cdot \lambda \cdot D} \left[ \frac{1}{2} x^2 - \frac{\cosh(kxd) - 1}{k^2 \cosh(k\rho)} \right] \qquad (11)$$

where $\Delta\psi$ represents the difference in CTE between the chiplets and the substrate in $°K^{-1}$, $\Delta T$ represents the thermal load. It should be noted that thermal load is different from the chip operating temperature (where $T_{peak} = 75°C$), as thermal load refers to the thermal cycling that occurs during the manufacturing process. We consider the thermal cycling range between $25°C - 105°C$. The coefficient $\tau$ is the thickness of the interposer, $\lambda$ is the thermal conductivity, $D$ is the stiffness factor of the interposer, and $k$ is the Youngs' modulus of the interposer. The variables $d$ and $\rho$ represent half the length of a chiplet and half the length of an interposer, respectively. As the placement, interposer size, chiplet dimension changes, the amount of warpage varies. For instance, placing embedded chiplets reduce the effective height $\tau$ of the interposer, directly reducing the warpage. Following existing work, we consider tolerable limit $\kappa_{max} = 150 \mu m$, which is the warpage constraint used in this paper [5]. We present a detailed warpage analysis in Section 4.3 (Fig. 10).

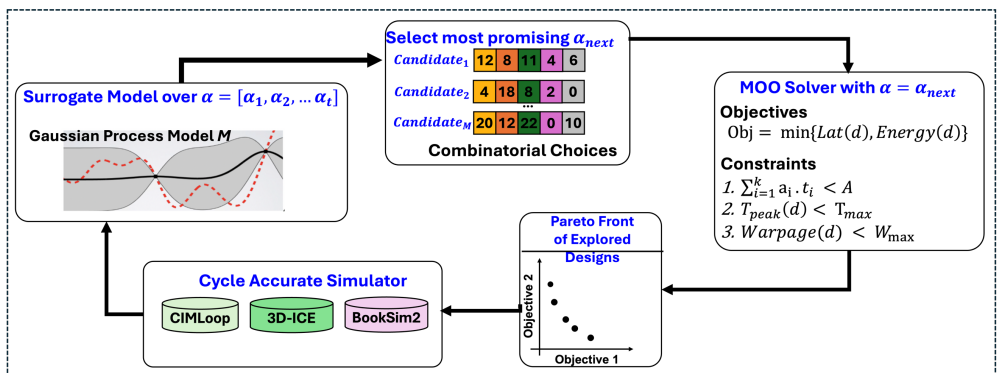

Figure 5: High-level overview of the proposed iterative design optimization framework to find pareto optimal designs to trade-off multiple objectives by meeting the package-level constraints.

### 3.3 Design Optimization Methodology

Our goal is to find the set of Pareto-optimal designs to trade-off multiple objectives, including performance, thermal, energy, and fabrication costs, subject to the area, peak temperature, and warpage constraints. There are two key challenges in solving this complex MOO problem. First, the combinatorial design space is *huge* and *conditional* in the sense that the problem of chiplet placement and mapping of neural layers to chiplets is conditioned on the selected composition of heterogeneous chiplets $\alpha = [\alpha_1, \alpha_2, ..., \alpha_t]$. Second, we need to perform expensive computational simulations to evaluate both performance and thermal objectives for any candidate design.

**High-level Overview of Methodology:** To address the first challenge of conditional design space exploration, we consider an alternating optimization approach that iterates between selecting a composition of heterogeneous chiplets $\alpha = [\alpha_1, \alpha_2, ..., \alpha_t]$ that meets the area constraint in the outer optimization step and solving the MOO problem for chiplet placement and neural layers to chiplet mapping in the inner optimization step by considering both temperature and warpage constraints. Fig. 5 shows the proposed optimization flow. In the inner optimization step, we only consider the candidate designs (chiplet placement and neural layer to chiplet mapping) that meet the warpage constraint during the design space exploration process, and we can employ any MOO solver (e.g., AMOSA or ML-driven solvers). Once we get a set of Pareto-optimal designs $D$ from the inner optimization step, we run detailed cycle-accurate simulations on these solutions to get absolute values for energy, performance, and temperature. Here, the solution quality is characterized by the energy-delay product (EDP) and the solutions that do not meet the temperature constraint are discarded. The EDP is a combined metric for performance and energy, defined as the product of system latency and total energy consumption. Hence, our goal is to find designs that minimize the EDP objective. Consequently, to evaluate the quality of a given composition of heterogeneous chiplets $\alpha = [\alpha_1, \alpha_2, ..., \alpha_t]$, we use the best EDP obtained from the inner optimization solver. We employ a Bayesian optimization approach for the outer optimization step because each call to the inner optimization step to get the best EDP is a computationally expensive procedure and our goal is to minimize the number of inner optimization calls.

**MOO Solver for a Given $\alpha = [\alpha_1, \alpha_2, ..., \alpha_t]$.** MOO methodologies based on Genetic Algorithms (NSGA-II) and Simulated Annealing (AMOSA) have been previously employed to handle multiple conflicting objectives effectively [40] [41]. These MOO methods consider a significantly smaller design space and converge slowly to the Pareto optimal design set. For example, the well-known AMOSA algorithm is based on simulated annealing. It needs to be annealed slowly to ensure a good solution, which does not scale well for large combinatorial spaces as in our case [40]. Prior work has shown that we can significantly improve the accuracy and speed of design optimization by utilizing ML procedures to guide the search. The key idea is to use the ML-driven knowledge gained from past explored designs to search large design spaces intelligently. Therefore, we employ MOOS algorithm as our solver [41]. MOOS performs adaptive design space exploration based on the principle of *optimism in the face of uncertainty*, which suggests exploration of the most favorable region of the design space based on the experience gained from past explorations. This data-driven approach not only guides the search towards better solutions and also reduces the runtime of MOOS to find acceptable solutions. Hence, we employ MOOS as the underlying MOO solver. To incorporate domain constraints into MOOS, we only consider valid perturbations (neighboring designs) such that the total system area is within the interposer area and the warplane constraint is satisfied and safely prune invalid designs during the design space exploration process.

Table 4: Cost comparison of TSVs between silicon and glass interposer.

| TSV cost overhead characterization | Silicon Interposer | Glass Interposer |
|---|---|---|
| TSV Count per Chiplet | 32 | 128 |
| TSV Area Increase Factor Area | 1x | 16x |
| TSV Vertical Depth | 150um | 150um |
| TSV Radius | 10-25um | 40um |
| TSV Cost (relative to interposer unit area) | 1x | 64x |

**Bayesian Optimization over $\alpha = [\alpha_1, \alpha_2, ..., \alpha_t]$.** BO is a derivative-free method to adaptively and efficiently search a given input space (search space of compositions of heterogeneous chiplets in our case) to optimize expensive-to-evaluate objective function (EDP of the best design for a given composition of heterogeneous chiplets). BO is an adaptive procedure to intelligently selects inputs for objective evaluation. The key ingredients of BO are: *1) a surrogate model* that captures our beliefs, based on past objective function evaluations, about the input-output relationship; and *2) an acquisition function* that scores each input according to the utility of querying the objective function on it next. We employ a Gaussian process (GP) surrogate model with Radial Basis Function kernel using training data in the form of $\alpha = [\alpha_1, \alpha_2, ..., \alpha_t]$ (input) and the best EDP from the inner optimization step (output). The decision of which input to evaluate next is made by maximizing the acquisition function. We only consider the $\alpha = [\alpha_1, \alpha_2, ..., \alpha_t]$ that meet the given area constraint during this selection step. We employ the expected improvement (EI) acquisition function due to its simplicity and effectiveness. Below, we describe the efficacy of our proposed design and optimization methodology comparing silicon- and glass interposer based chiplet systems with different power, performance, area, and cost (PPAC) tradeoffs.

## 4 EXPERIMENTAL RESULTS

In this section, we present a thorough performance analysis of the glass interposer-based 2.5D architectures and compare their performance and thermal efficiency with respect to silicon interposer-based 2.5D architectures. For the performance evaluation, we consider all the NoI architectures mentioned in Table 1.

### 4.1 Experimental Setup

To demonstrate the performance and energy efficiency of the proposed multi-chiplet glass interposer systems, we consider two different configurations with a silicon interposer area limit of $400 \ mm^2$ and $864 \ mm^2$, respectively [39]. Table 1 shows different device types, storage capacity, frequency of operation, area footprint, $TOPS$ and $Energy/MAC$ (fJ/MAC) for different PIM chiplets considered in this work. Following the existing PIM implementations, we consider the PIM chiplet sizes to be within 10mm$^2$ [3] [4] [8]. However, it should be noted that the area footprint of chiplets can even be in hundreds of millimeters as has been demonstrated in industrial explorations [3] [42] [43]. As each PIM chiplet area may vary, the total number of chiplets, depending on the chosen types within the given interposer area, will also vary. Weights for each set of incoming DNNs are assumed to be loaded through the DRAM stacks on the periphery of the accelerator, as shown in Fig. 1. In this work, we focus on evaluating the performance of the glass interposer compared to the silicon interposer using the proposed design framework. We consider various NoI architectures proposed so far as the communication backbone when evaluating different silicon and glass interposer-based systems [4] [8] [9] [10]. Hence, NoI optimization is orthogonal to this work. To model chiplet-based systems with varying architectures, we develop a custom *tool flow* to characterize the overall performance and thermal profile by considering the circuit- and system-level configurations. Fig. 6 illustrates the chosen tool flow to evaluate the MOO outputs accurately. We integrate circuit (HSPICE), system (CIM-Loop), network (BookSim2), and thermal (3D-ICE) simulators to determine the overall performance, energy consumption, accuracy, and thermal profile of each multi-chiplet architecture under consideration [44] [45] [46]. It should be noted that other tools, such as NeuroSim, MFIT, and PACT, can also be incorporated into this evaluation framework while maintaining similar accuracy [47] [48] [49]. We employ CIM-Loop to model diverse PIM configurations, including its microarchitectural configurations such as row/column drivers, ADCs, memory cells, and digital components including adders, and accumulators [50]. CIM-Loop enables accurate data value dependent modeling, necessary for ReRAM-based chiplets. Each DNN task is partitioned and mapped onto the multi-chiplet system. The inter-chiplet traffic is generated by the expected activation flow among the neural layers. In our NoI modeling, we maintain the same operational frequency for all PIM chiplets. However, the inter-chiplet communication delay varies with the wire length. We model the inter-chiplet delay using HSPICE following the model proposed in the existing literature [51]. We consider energy-per-bit, router frequency, and number of channels consistent with Nvidia GRS and UCIe protocol [12] [38]. In the case of the interposer, the communication is through a die-to-die interface through the interconnect wire. We then characterize the NoI performance using BookSim2. The number of chiplets, their placement, and the NoI architecture are the input to the BookSim2 simulator along with the inter-chiplet traffic traces. Finally, we employ 3D-ICE to find the thermal profile of the system [45] [46].

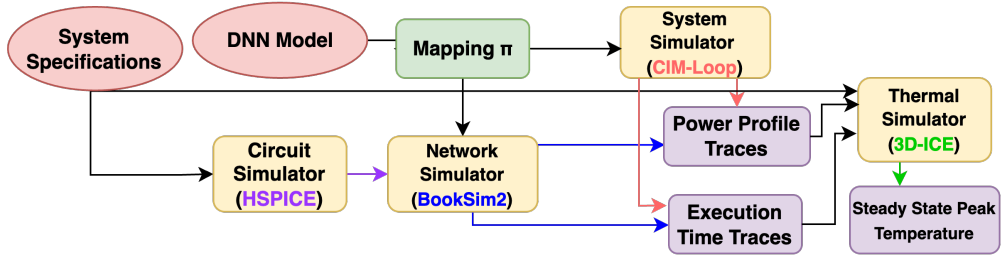

Figure 6: EDA tool flow for evaluating heterogenous multi-chiplet systems.

Table 5: List of DNN workloads evaluated with their number of parameters (in millions).

| Workload | CNN tasks | Total number of parameters |
|---|---|---|
| WL1 | ResNet18, ResNet34, ResNet152 | 177M |
| WL2 | ResNet34, VGG16, ResNet18 | 171M |
| WL3 | ResNet34, VGG19 | 166M |
| WL4 | VGG19, MobileNetV2, DenseNet169 | 161M |

### 4.2 Baseline System Design and DNN Application Workloads

We evaluate the performance and energy efficiency using multiple DNN workloads with up to a billion parameters. Table 5 shows the considered DNN inferencing workloads WL1-WL4 on the ImageNet dataset with their number of parameters in millions. For an interposer area $A$ (within 400mm$^2$), we consider an ensemble of concurrent DNN inference tasks representing a server-scale environment. In each workload (WL), we consider different numbers and types of DNNs running in parallel. Note that, the DNNs considered in our evaluations consist of linear (VGG), residual (ResNet), as well as dense (DenseNet) connections. Moreover, all the DNNs comprise of both fully connected and convolution layers. Each workload is executed at an 8-bit precision. In these evaluations, we determine the execution time and energy consumption on chiplet systems for each workload. We evaluate the effectiveness of the proposed glass interposer system with respect to the silicon interposer counterparts using the previously mentioned types of chiplets: *Standard, Shared, Adder, Accumulator*, and $ADC_{Less}$. We consider several scenarios to evaluate the efficacy of glass interposer systems in comparison to silicon interposer counterparts considering either the same chiplet configuration or fabrication cost. As glass interposer is cheaper, we can increase the interposer area size by a certain degree to increases the system TOPS.

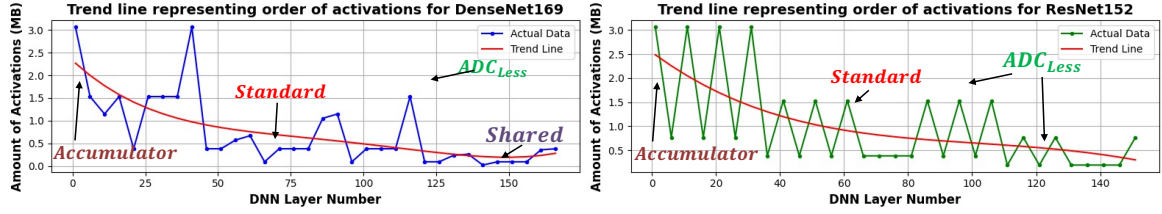

Figure 7: Illustrative example showing the number of activations for different neural layers of DenseNet169 and ResNet152. The number of activations reduce for later layers. The ideal PIM configuration changes as shown depending on the neural layer requirements.

For the given interposer size, we choose different chiplet configurations using the MOO procedure to jointly optimize the performance and energy consumption. Subsequently, we compare the performance of silicon and glass interposer-based systems. As the individual chiplet size increases, the number of allowable chiplets that could be placed on the interposer decreases. Hence, the total number of chiplets depending on the chosen type varies. For instance, if we execute an ensemble of DNNs (WL1 to WL4) with only $ADC_{Less}$ chiplets which are smaller than *Shared*, we can integrate 110 chiplets. Using the MOO procedure, 82 heterogenous chiplets are employed with the following breakup [24, 28, 0, 18, 12] where each element value in the list corresponds to the total number of chiplets from *Standard*, *Shared*, *Adder*, *Accumulator*, and $ADC_{Less}$ types of PIM configuration. It can be observed that *Accumulator* offers superior storage capacity and higher TOPS compared to *Adder* (refer Table 1). Hence, no *Adder*-based chiplet is chosen by the design optimization framework for accelerating DNN inference tasks.

To justify the merit of the proposed design framework, we need to consider the flow of activations and weight storage characteristics of the DNN workloads under consideration. Fig. 7 shows an illustrative example of the number of activations for different neural layers of DenseNet169 and ResNet152. The arrows in Fig. 7 highlight the preferred PIM configuration depending on the layer characteristics. In both DNNs, due to the max-pool layers, the number of activations reduces for the later layers. However, the weight storage of each of these later layers may increase. For example, for ResNet152, the weight storage requirement of the 2[nd] layer is 25KB. However, the weight storage requirement of the 100[th] layer increases to 600KB. A similar trend is observed for other DNNs as well. MOO procedure effectively chooses the PIM configuration, which is ideal for different layer(s) by considering their individual characteristics. As illustrated in Fig. 7, as the number of activations reduce for the later layers, lower storage and energy consumption chiplets such as $ADC_{Less}$ can be employed for energy efficiency. For later layers, depending on the DNN type, *Shared* type of chiplets might be preferred. Similarly, *Accumulator* PIM configuration is chosen for initial layers since higher compute is preferred to accelerate the inference pass. Table 6 shows a qualitative comparison based on application requirements and preferred type of PIM chiplet. As the energy efficiency, TOPS, thermal sensitivity, power profile, and storage vary, the optimal choice is application dependent. Hence, the multi-chiplet heterogeneous system design inherently requires design space exploration.

Table 6: Qualitative comparison between different types of PIM chiplet architectures.

|  | Standard | Shared | Adder | Accumulator | ADC$_{Less}$ |
|---|---|---|---|---|---|
| **Energy Efficiency** | Lower | High | Medium | High | Highest |
| **Performance** | Medium | Medium | Low | High | Low |
| **Storage** | High | High | Low | High | Low |
| **Power Density** | High | Low | Medium | Medium | Low |
| **Thermal resilience** | Bad | Good | Good | Bad | Good |
| **Best Use Case** | General purpose AI acceleration | Memory-intensive operations | Energy-constrained Applications | High-compute parallel applications and DNN Training | Ultra-low power applications/ Binary neural networks |

### 4.3 Overall Performance Evaluation

In this section, we first present a thorough performance evaluation of different multi-chiplet architectures executing concurrent DNN inference tasks (as server-scale workloads) with hundreds of millions of parameters, as shown in Table 5. We first consider different NoI designs and study their impact on the optimized heterogeneous multi-chiplet systems.

#### A. Impact of NoI Architectures over Heterogenous Systems

In this subsection, we characterize the impact of the NoI for power, performance, and thermal trade-offs when executing WL1-WL4. We consider this evaluation to characterize the impact of NoI on achievable performance. We consider recently proposed NoI architectures, including Kite, Mesh, HexaMesh, and Floret [8] [10] [12] [13] [33]. Fig. 8 shows the Kite, HexaMesh, Mesh, and Floret topologies for a system with 36 chiplets as demonstrative examples. It can be observed among the considered NoI architectures, Floret has the least number of inter-chiplet links and HexaMesh has the highest. Between communicating pair of chiplets, the average hop count $H_{avg}$ is defined as:

$$H_{avg} = \frac{\sum_{j=1}^{N} \sum_{i=1}^{N} F_{ij} \times H_{ij}}{\sum_{j=1}^{N} \sum_{i=1}^{N} F_{ij}} \tag{12}$$

where $F_{ij}$ and $H_{ij}$ is the traffic volume and the hop count between two communicating chiplets. As the traffic pattern $F_{ij}$ changes for different DNN workloads, the $H_{avg}$ for different topologies changes. Hence, instead of mentioning the static hop distance, we need to capture the overall latency of each NoI. This is because latency is influenced by the mapping of layers onto the topology, which determines traffic patterns, contention points, and the specific links utilized during a data transfer. Moreover, the traffic is heavily irregular in case of DNN workloads. Therefore, two topologies with similar average hop distances can exhibit different latency behaviors depending on how the workload is distributed across the network. It should also be noted that any k-ary n-cube based topologies like Kite, Mesh, HexaMesh, among others can be degenerated to an equivalent SFC-based architecture to improve their performance and energy efficiency by reducing their average hop count for different DNNs [52].

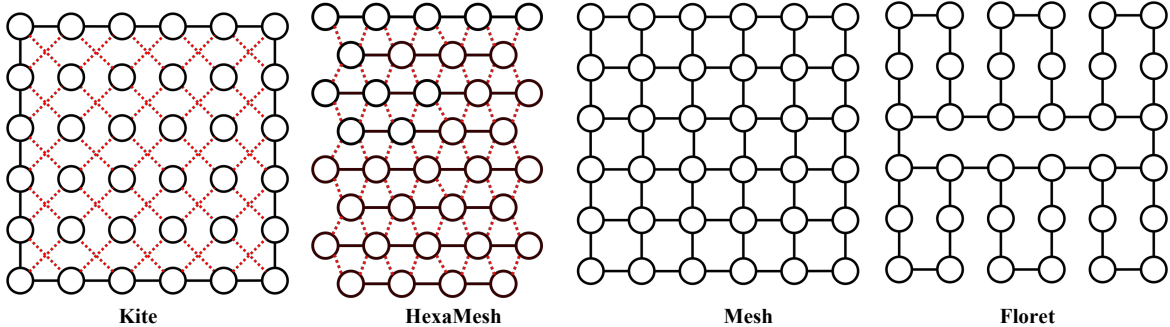

Figure 8: Different topology with their link configuration.

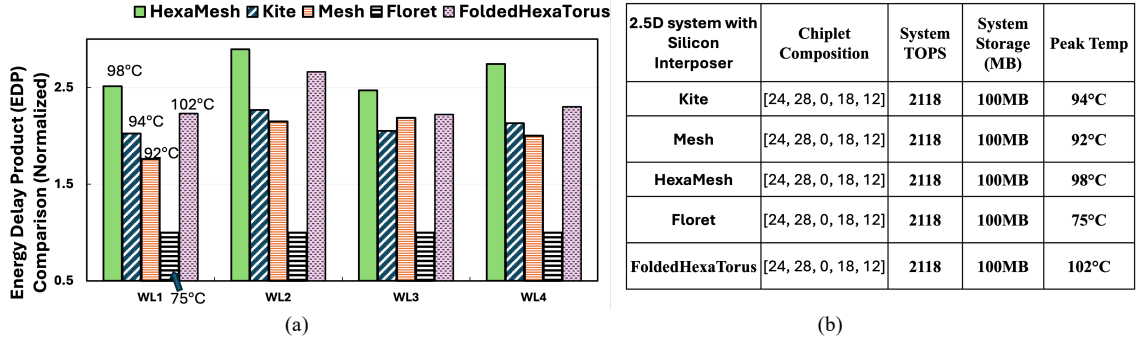

Figure 9: (a) EDP comparison when executing WL1-WL4 on different NoI architectures including HexaMesh, Kite, Mesh, FoldedHexaTorus, and Floret; (b): Chiplet composition, system TOPS, system storage, and peak temperature when executing DNN models with different NoIs.

Fig. 9(a) shows the energy-delay-product (EDP) comparison with chiplets interconnected using different NoI architectures on a silicon 2.5D system. It can be observed that Floret-NoI improves EDP on average by 2.64×, 2.1×, 2.47×, and 2× in comparison to HexaMesh, Kite, FoldedHexaTorus, and Mesh-based chiplet systems, respectively. Moreover, as the number of router ports for each NoI varies, the overall power profile for the system varies. HexaMesh and FoldedHexaTorus, with six router ports, have the highest power consumption, followed by Kite and Mesh with four routers, with the lowest power with Floret, with most chiplets only having two router ports. Hence, the peak temperature follows the same ordering with the highest from FoldedHexaTorus ($T_{peak}$=102°C), HexaMesh ($T_{peak}$=98°C), followed by Kite ($T_{peak}$=94°C), Mesh ($T_{peak}$= 92° C). Floret, due to its dataflow awareness and reduced number of links and associated router ports, has the lowest peak temperature of 75°C. Among all the NoI architectures, Floret is the only NoI that satisfies the thermal constraint. Hence, for further investigation to compare silicon and glass interposer systems, we employ Floret as the communication backbone for all comparative evaluations unless otherwise stated. We refer to silicon interposer and glass interposer-based systems as Silicon_2.5D and Glass_2.5D, respectively.

### B. Silicon vs. Glass Interposer-based 2.5D System

In this subsection, we present a thorough performance evaluation with different chiplet configurations previously discussed. Specifically, we compare and contrast glass interposer-based heterogenous systems of each type (Standard, Shared, Adder, Accumulator, and $ADC_{Less}$) with respect to their silicon-based 2.5D counterparts. We first characterize the impact of the thermal variation on the DNN accuracy. All the performance and energy consumption plots are normalized with respect to Glass_2.5D. Fig. 10(a) shows the performance comparison between Silicon_2.5D and Glass_2.5D systems executing WL1-WL4 along with the peak system temperature. We show the corresponding breakdown in overall execution time between compute and communication time. Glass_2.5D chiplet system has higher bus width, lower RDL capacitance, and hence lower energy per bit with a higher frequency of operation during communication phase (2GHz vs. 1.15GHz) [5]. Hence, on average, it has about 1.3x lower end-to-end latency in comparison to its silicon_2.5D counterpart. However, the peak temperature is much higher in comparison to Silicon_2.5D system. With a higher peak temperature, ReRAM-based chiplets exhibit conductance variation in the weight values which leads to accuracy loss. Hence, we need to constrain the system within an allowable peak temperature limit of 75°C. This is because conductance drift in ReRAM-based PIM chiplets can lead to degradation in DNN inference accuracy, especially under elevated temperatures. As the temperature increases, the OFF-state conductance value of the ReRAM cell ($G_{Off}$) increases exponentially with the temperature [27] [53]. This degradation is workload-dependent, as the weight distributions and sensitivity to perturbations vary across different DNN architectures.

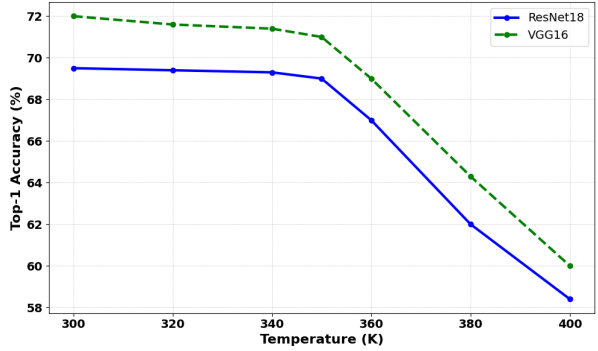

Figure 10: Impact of thermal induced conductance drift as the neural network changes on Glass_2.5D system.

To analyze this, we employ temperature-induced weight drift model using the relationship $G_{Off} \propto \exp[\eta T]$, where T is the operating temperature in Kelvin and η is the temperature coefficient [53]. Following prior work, the weight variation is modeled as a normal distribution $\Delta w \sim N(0, \sigma^2)$ where σ increases with temperature [54]. Weights and activations of the neural layers are stored in ReRAM cells as conductance states. The change in OFF-state conductance $G_{Off}$ affects the weight values that in turn introduces errors in the output produced by the ReRAM crossbar array. It should be noted that not all neural layers are mapped to ReRAM-based chiplets. the heterogeneous architectures in Silicon_2.5D and Glass_2.5D system include both SRAM and ReRAM chiplets, and the impact of conductance drift is a function of the neural layer-to-chiplet mapping. Only the layers mapped to the ReRAM chiplets are affected by the temperature-induced conductance variation and hence suffer accuracy loss. Furthermore, due to workload-dependent mappings, different chiplets reach different peak temperatures, leading to non-uniform accuracy degradation across neural networks. Hence, the overall accuracy of the model is affected.

As an example, Fig. 10 shows the Top 1 accuracy degradation for VGG16 and ResNet18 as an example when the peak temperature varies on Glass_2.5D system. Notably, accuracy begins to degrade significantly beyond 75°C (348 K), depending on which layers are mapped to thermally sensitive ReRAM regions. This underscores the importance of designing thermally aware heterogeneous architectures. It can be observed that VGG16 has higher accuracy degradation percentage in comparison to ResNet18. VGG16 uses ReLU activations in convolution layers and small weight changes can push activations into saturation regions. This can degrade the entire feature maps, causing dramatic accuracy drops. Hence, it is necessary to maintain the peak temperature within 75 °C.

It can be observed from Fig. 11(a) that the Glass_2.5D systems have much higher $T_{peak} = 86°C$ with the same chiplet configuration as Silicon_2.5D. This necessitates changing the chiplet configuration to ensure thermal constraint is met. The MOO optimization method proposed the following system configuration: [2, 27, 2, 27, 15] with TOPS = 1813; System

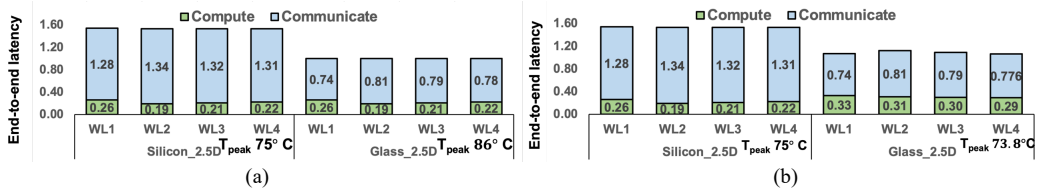

Figure 11: Latency comparison when executing WL1-WL4 on (a) Silicon_2.5D & Glass_2.5D system with same configuration;(b) Silicon_2.5D & (b) Glass_2.5D system optimized for peak temperature within allowable limits.

storage = 98MB, and $T_{peak} = 73.8°C$ within the interposer area limit $A$. This leads to about 15% lower system TOPS, but the peak temperature is within the tolerable limit of 75°C. Moreover, as the DNN workloads are fixed and weights need to be mapped, it is imperative that the chosen chiplet configuration supports high storage comparable to the baseline design to fully map the DNN workloads. This is because the compute-centric chiplets (with higher power density) are replaced with comparably lower power chiplets, which may have lower storage, to balance the peak system temperature. Fig. 11(b) shows the performance impact in reducing the peak temperature of the Glass_2.5D system in the same system size of 400mm². It can be observed that as the total system TOPS have reduced by 15%, there is a performance degradation for each workload as the average compute latency increases by ~39%. This is because even though the system TOPS reduces by about 15%, the corresponding neural layer execution time increases in different proportions depending on the neural architecture, which leads to cascaded effects in the overall computation time. However, communication is still the bottleneck and a major contributor to the overall execution time. Hence, with reduced system compute capability for ensuring thermal feasibility, Glass_2.5D has higher performance in comparison to Silicon2.5D. We can observe that Glass_2.5D system achieves on an average 1.41x improvement over the Silicon_2.5D counterpart, with up to 40% power reduction. To further analyze the thermal behavior of Glass_2.5D systems, we consider the impact of the interposer's lower heat transfer coefficient on overall system performance. Due to reduced thermal conductivity, glass interposers limit heat dissipation, resulting in elevated peak temperatures under identical chiplet configurations. We conduct a sensitivity analysis comparing Glass_2.5D and Silicon_2.5D systems across two representative chiplet configurations, using Floret as the NoI backbone. Fig. 12(a) presents the energy-delay product (EDP) for both systems under configuration $\alpha_{optimized}$ = [2, 27, 2, 27, 15], where all systems maintain identical chiplet compositions, i.e., identical system storage, and TOPS. This is the configuration the optimizer proposes for Glass_2.5D system within the thermal constraint of 75°C. Despite thermal constraints, Glass_2.5D achieves an average 2.06× lower EDP than its Silicon_2.5D counterpart. Fig. 12(b) extends the comparison using the original Silicon_2.5D configuration $\alpha_{initial}$ = [24, 28, 0, 18, 12]. This is the initial configuration used for Silicon_2.5D system. Here, Glass_2.5D achieves an average of 1.5× lower EDP in comparison to Silicon_2.5D counterpart. On the contrary, the maximum temperature in the Glass_2.5D system has a maximum overshoot of ~ 8.6°C over the thermal limit of 75°C when executing WL4. Silicon_2.5D remains within the thermal limit of 75°C. This result highlight that although glass interposers necessitate thermal-aware design space adaptations, they can offer substantial performance advantages over silicon under the interposer area limit.

### C. Scalability of Glass Interposer-based Systems

Integrating chiplets on the glass interposer presents a trade-off between performance gain and thermal bottleneck. However, as the system size scales, there is a 'reticle' or 'limit' on the interposer size due to the amount of deformity experienced by the interposer (refer eq. (11)). Fig. 13(a) shows the comparison of warpage in Silicon_2.5D and Glass_2.5D system. It can be observed that the amount of warpage is much higher for the Glass_2.5D system in comparison to the silicon counterpart. Hence, there is a limit on interposer dimension, as shown in Fig. 13(a) as 20mm. It should be noted that the 20mm is for the physical dimension of the interposer and not the link length. The inter-router link lengths depend on the NoI configuration. The maximum allowable interposer size for a standard Glass_2.5D system is 20X20 or $400 mm^2$ beyond which a Glass_2.5D system is unrealizable. Hence, necessary optimizations are required to enable a larger glass interposer-based system. Embedding functional components within the glass interposer has been explored as a technique to potentially reduce the amount of warpage. In this section, we characterize the impact of embedding chiplets to enable architecture and package co-optimization to enable glass interposer-based chiplet systems.

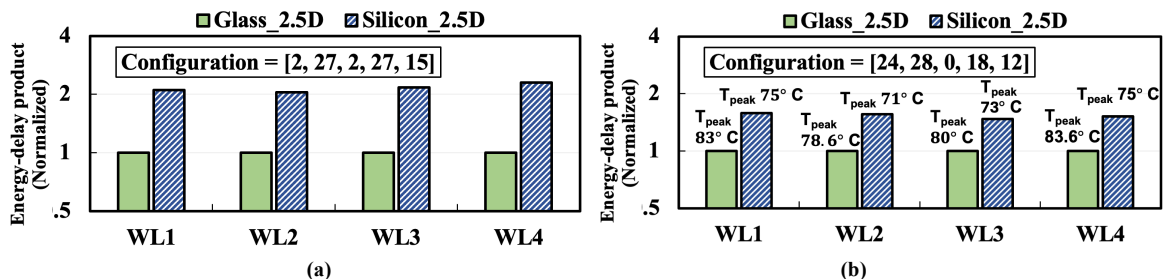

Figure 12: EDP comparison when executing WL1-WL4 on (a) Silicon2.5D with the downsized configuration as Glass_2.5D; (b)Silicon_2.5D, Glass_2.5D system with same initial Silicon_2.5D chiplet composition.

Fig. 13(b) shows the warpage comparison between the Silicon_2.5D system and Glass_2.5D system with embedded chiplets. We assume to embed chiplets farthest from the center of the interposer to counter the warpage the most in this experiment. It can be inferred that as the number of embedded chiplets increases, the effective height of the interposer reduces. The stiffness of the interposer is proportional to its height. Hence, the overall interposer stiffness increases and warpage reduces. With about 10% embedding of chiplets, we have the same amount of warpage with respect to a Silicon_2.5D system. This is imperative as this enables us to create large-scale Glass_2.5D systems. However, due to embedding of the chiplets, the power density of the system and hence the peak temperature increases. In comparison to a Glass_2.5D system with no embedded chiplets, a system with about 10% of area occupied by embedded chiplets can lead to 15° higher peak temperature. Glass_2.5D_Embed enables high-performance systems within the thermal limit and warpage. As the PIM chiplet configuration varies, the achievable performance, energy consumption, amount of warpage, and fabrication costs vary. In the next section, we aim to highlight the PPAC trade-offs between Silicon_2.5D, Glass_2.5D, and Glass_2.5D_Embed (with embedded chiplets), as system size scales from 400 $mm^2$ and to 864 $mm^2$. In this work, we only embed SRAM type of chiplets, and no ReRAM-based chiplets are embedded.

### D. PPAC Trade-off Analysis

In this section, we compare the PPAC trade-offs across four system configurations: standard Silicon_2.5D, Silicon_2.5D with active interposer support (Silicon_2.5D_Embed_Router), standard Glass_2.5D, and Glass_2.5D with embedded chiplets (Glass_2.5D_Embed). We begin by contrasting the standard Silicon_2.5D system with the active interposer variant (Silicon_2.5D_Embed_Routers), where NoI router logic is embedded into the interposer. Please note that Glass_2.5D system is unrealizable at larger interposer size of 864 $mm^2$ due to excessive warpage. Hence, only Silicon_2.5D and Glass_2.5D_Embed are characterized at a larger system size.

**Existing Active Interposer system:** Modern packaging solutions such as Intel EMIB, TSMC InFO-oS, and IntAct enable such lightweight logic embedding but not full chiplet integration [4] [17] [18]. In our experiment, the router area is reclaimed to add more compute chiplets on the surface, keeping total area constant. However, these designs incorporate only lightweight logic blocks such as routers, buffers, or small interconnect units but not full chiplets. We undertook an experiment to characterize performance for an active interposer system that only embeds the NoI routers. We use the area saved by embedding the routers to introduce more computational units (chiplets) as surface chiplets on the 2.5D based system. As the NoI topology varies, the number of ports and associated link differs. Hence the total available router area depending on the topology varies. Table 7 shows the percentage of chiplet area corresponding to router depending on the NoI topology along with the number of operations (in tera-operations per second or TOPS) before embedding router and after embedding router for each NoI.

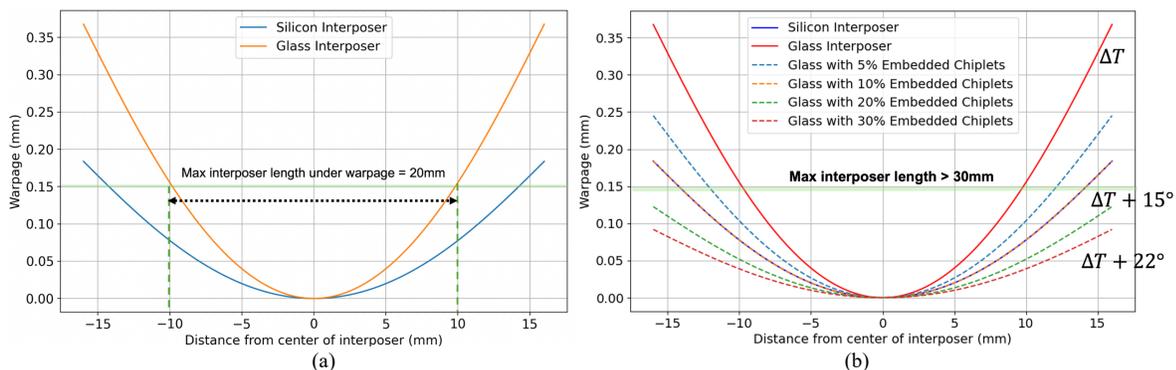

Figure 13: Warpage comparison between (a) Silicon_2.5D & Glass_2.5D system; (b) Silicon_2.5D & Glass_2.5D_Embed

Table 7: Router area percentage along with initial and final TOPS.

| | Router Area Percentage | Initial TOPS | Final TOPS |
|---|---|---|---|
| Kite | 4.54% | 2118 | 2148 |
| Mesh | 4.54% | 2118 | 2148 |
| HexaMesh | 6.66% | 2118 | 2154 |
| Floret | 2.42% | 2118 | 2136 |

Fig. 14 shows the impact on the overall EDP by embedding only the routers and using the available silicon area for computation. In this experiment, we consider Floret and HexaMesh as the two NoI backbones. These two NoIs represent the two extreme cases in terms of the overall router area. HexaMesh, due to larger number of router ports has the highest router area. Floret, with its space filling curve-based path has mostly two ports for each router. Hence, Floret introduces the least router are overhead. It can be observed in the Fig. 14 that while the overall TOPS increases by introducing active components, the overall performance improvement is limited. This is because the maximum TOPS increase (for HexaMesh) is only 1.69% of the initial value. Moreover, the realizable performance improvement is dependent on computation, layer weights, number of MAC operations, and the on-chip communication pattern. For brevity, we characterize the energy delay product for WL1 and WL2 only. It can be observed that in case of HexaMesh, the EDP reductions by embedding routers in the interposer system for WL1 and WL2 are ~0.5% and 2.3% respectively. Similarly, for Floret NoI, we observe a 1.4% improvement when executing WL2. A similar trend is observed for all other DNN workloads. The overall compute capacity only increases by 0.84-1.69% which does not yield huge performance benefits.

In the context of Glass_2.5D_Embed system, our use of 'embedded chiplets' refers specifically to standalone PIM-based units. Typically, they are not embedded directly into silicon interposer due to additional cost, yield, thermal, reliability issues and power delivery limitations as demonstrated in existing literature [1]. Below we compare PPAC trade-offs under two primary constraints: (i) fixed chiplet configuration and (ii) fixed fabrication cost, with an emphasis on performance, power, area, and thermal feasibility at increased interposer sizes for Silicon_2.5D and Glass_2.5D_Embed.

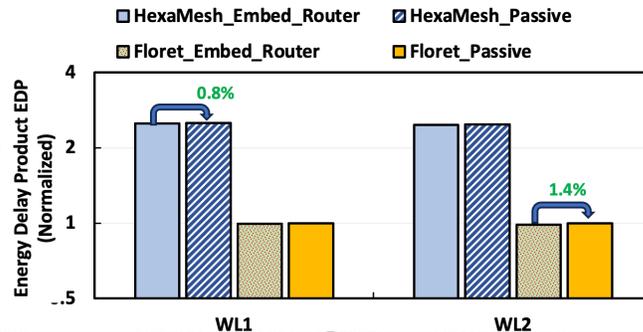

Figure 14: EDP comparison when executing WL1 and WL2 comparing Floret and HexaMesh NoI with active interposer.

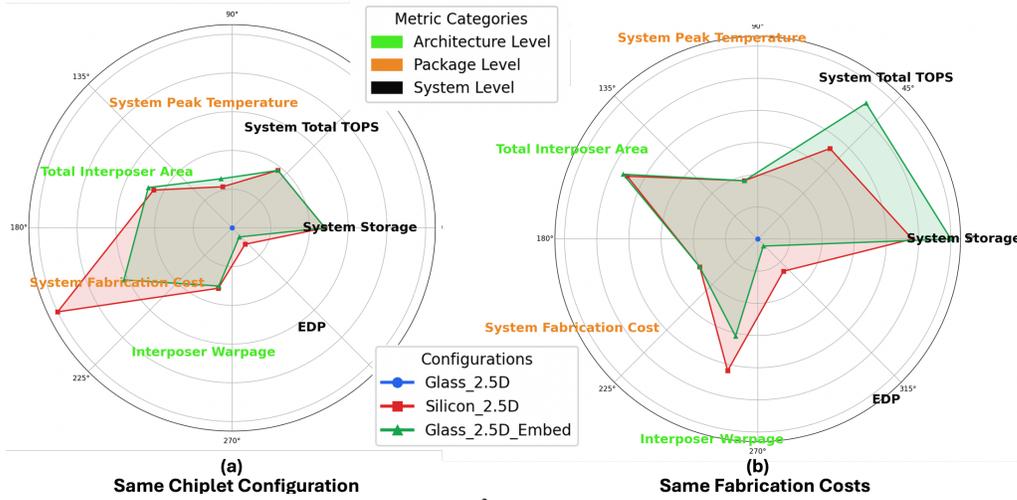

Figure 15: PPAC trade-off with larger interposer area of 864mm$^2$ employing (a) Same Chiplet Configuration; (b) Same Fabrication Cost.

**Under the same chiplet configuration:** As the chiplet configuration is the same, we are disintegrating the chiplets into surface and embedded chiplets. Every embedded chiplet decreases the warpage but increases the thermal overhead. Hence, the MOO method explores different combinatorial configurations of surface and embedded chiplets. Fig. 15(a) shows the PPAC trade-off comparison between silicon and Glass_2.5D_Embed system when scaling interposer area. It can be observed that the System TOPS and System Storage are identical for both Silicon_2.5D and Glass_2.5D systems as they have the same chiplet configurations. It can be observed that Glass_2.5D_Embed enables 45% lower EDP on average in comparison to the Silicon2.5D system. Moreover, the fabrication cost is lower by 3.1x in comparison to the Silicon_2.5D system. When the system size scales, the EDP reduction is about 64% in comparison to the Silicon_2.5D counterpart. This is due to faster communication enabled with Glass_2.5D. It can be observed that for enabling the same configuration, both Glass_2.5D_Embed have 6° higher peak temperatures.

**Under the same fabrication cost:** We can constraint the fabrication cost and compare the PPAC trade-offs between silicon and Glass_2.5D systems. Following (12), between Silicon and Glass interposer-based system, about 180mm$^2$ of extra embeddable area is available. The proposed MOO method enables configuration which maximizes the performance, ensuring the system is within the thermal and warpage limits. Following is the optimized chiplet configuration: Surface: [12, 28, 0, 18, 48]; Embedded Chiplets: [0,12,12,0,12]. Fig. 15(b) shows the PPAC trade-off comparison between silicon and Glass_2.5D when scaling the interposer area. It can be observed that as the fabrication cost is the same between silicon and Glass_2.5D_Embed systems, MOO effectively chooses different chiplet configurations to balance performance, thermal, warpage metrics, within the thermal constraints. Hence, the system TOPS and system Storage is different between Silicon_2.5D and Glass_2.5D systems. It can be observed that the EDP reduction from a smaller system size to a larger system size is more pronounced with 1.8x reduction. Moreover, the peak temperature is within the allowable limit of $T_{max}$. The amount of warpage is reduced by a factor of 20% in comparison to a Silicon_2.5D counterpart. By embedding more chiplets, Glass_2.5D_Embed enables higher on-chip storage and higher TOPS by balancing resources within the given thermal and logic area budget, mitigating the warpage issue for glass interposer-based 2.5D multi-chiplet architectures.

*E.* Extending Glass_2.5D systems for accelerating large language models:

To extend the applicability of the Glass_2.5D framework, we consider different LLM workloads for a fixed input sequence length of 256 tokens with up to 1.5 billion parameters. Fig. 16(a) shows the considered LLM models and their number of parameters (within $A_{int} \leq 400$ mm$^2$) considered in different PIM configurations.

LLMs primarily comprise an encoder and a decoder stack with identical computational structures [55] [56] [57]. The heart of the transformer model is the multi-head attention (MHA) kernel, which comprises of Key (K), Query (Q), and Value (V) computation, which are further processed to compute the Score ($S = \text{Softmax}(\frac{Q*K^T}{\text{sqrt}(d_k)})$) where $d_k$ is the dimension of the Key vector. The score represents the probability of attending to each key and must be computed continuously for given input for each head in each encoder and decoder, respectively. Hence, the attention kernel requires a massive number of rewrites during these computations due to large amount of data computation. In contrast to CNNs, attention mechanisms exhibit highly irregular traffic patterns, characterized by numerous source-to-destination communication pairs and non-sequential data dependencies. This irregularity arises from the all-to-all nature of attention computations, where each token may attend to every other token, leading to scattered data transfers. To accommodate this behavior, we designed an application-specific irregular NoI topology specifically for the LLM applications consistent with existing literature [9] [58]. After the attention mechanism, every LLM has a feed-forward (FF) network. The FF network consists of two consecutive fully connected (FC) layers, which are large static hidden layers. Like DNN models, the fully connected layers have fixed sizes and sparser weight vectors than the attention layer.

It should be noted that due to the write endurance concern, we do not consider any ReRAM-only architecture [55]. To illustrate, consider the DPT-Large model processing a token length of 1024, where each attention head is mapped to a unique ReRAM chiplet. We need $\sim 5 \times 10^4$ rewrite operations to ReRAM cells. Notably, the number of necessary rewrites increases with the sequence length due to dynamic matrix multiplications. Consequently, the MHA computation on ReRAMs can quickly approach the ReRAM endurance limit (rewrites exceeding $10^6$-$10^9$).

Contrarily, the FF network computation on ReRAM is independent of the input sequence length. Due to the fixed and limited number of updates, we can employ ReRAM to exclusively accelerating the FF network. This is a merit of employing heterogeneous chiplets for different computational kernels. In summary, the nature of the computational kernels in the transformer model necessitates a heterogeneous architecture for executing end-to-end LLMs.

In the case of multi-chiplet LLM accelerator design, Silicon_2.5D has the following heterogeneous chiplet breakup: [16, 16, 18, 24, 10] where each element value in the list corresponds to the total number of chiplets from Standard, Shared, Adder, Accumulator, and ADC$_{Less}$ types of PIM configuration. Glass_2.5D system has following chiplet breakup: [6, 24, 4, 26, 16]. Fig. 16(b) shows the EDP of various multi-chiplet architectures executing different LLM workloads. Each plot is normalized with respect to the EDP of the corresponding Silicon_2.5D system. It can be observed from Fig. 16(b) that Glass_2.5D based multi-chiplet architecture consistently outperforms Silicon_2.5D with up to 67%, 34%, and 55% reduction in EDP executing WL6, WL7, and WL8 respectively. Here, the peak temperature is 72.3°C when executing WL6, which is well within the limit for any possible thermal-induced accuracy loss when executing LLM models.

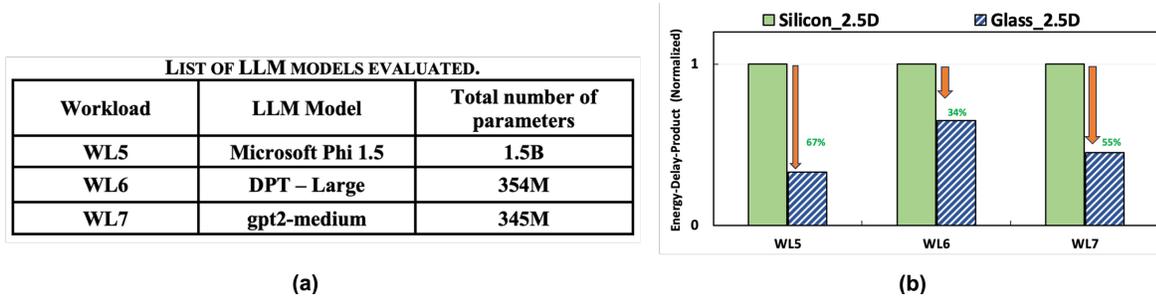

| LIST OF LLM MODELS EVALUATED. | | |
|---|---|---|
| **Workload** | **LLM Model** | **Total number of parameters** |
| WL5 | Microsoft Phi 1.5 | 1.5B |
| WL6 | DPT – Large | 354M |
| WL7 | gpt2-medium | 345M |

(a)          (b)

Figure 16: (a) LLM workloads considered when being executed on Glass_2.5D system; (b) Energy-delay-product (EDP) comparison for executing LLMs between Silicon_2.5D and Glass_2.5D systems.

## 5 CONCLUSION

This paper addressed the challenges in designing a multi-chiplet system using glass interposer. Glass interposer allows the scaling of system performance with lower energy per bit. However, thermal constraints and warpage limit the interposer size. We proposed a general architecture packaging co-optimization framework to enable suitable disintegration with a combination of surface and embedded chiplets. The glass interposer-based architectures have better PPAC trade-offs compared to silicon interposer-based architectures for relatively larger system sizes (wafer-scale chips). Our evaluations on server-scale deep neural network workloads demonstrate that glass interposer-based chiplet systems achieve up to 2.1x performance improvement, 40% energy reduction, with lower fabrication cost compared to silicon 2.5D counterparts.